\documentclass[3p,10pt]{elsarticle} 		
\journal{arXiv.org}

\usepackage{siunitx}		   	    							
\usepackage{IEEEtrantools} 								
\newcommand{\code}[1]{\small{\texttt{#1}}\normalsize}	
\newcommand*\chem[1]{\ensuremath{\mathrm{#1}}}			

\usepackage{subfig}										
\usepackage[hyphens]{url} 								
\usepackage[latin1]{inputenc} 							
\setcitestyle{sort&compress}								
\usepackage[modulo]{lineno}								
\usepackage[doublespacing]{setspace} 					

\begin{document}
\sloppy 													

\begin{frontmatter}

\title{Dosimetric evidence confirms computational model of magnetic field induced dose distortions of therapeutic proton beams}

\author[oncoray,hzdr-ior]{Sonja M. Schellhammer\corref{cor1}} \ead{s.schellhammer@hzdr.de}
\author[oncoray,hzdr-ior]{Sebastian Gantz}
\author[oncoray,hzdr-ior,dktk]{Armin L\"uhr}
\author[wol-uni,wol-hos]{Bradley M. Oborn}
\author[hzdr-iorp]{Michael Bussmann}
\author[oncoray,hzdr-ior,str]{Aswin L. Hoffmann}

\address[oncoray]{OncoRay - National Center for Radiation Research in Oncology, Faculty of Medicine and University Hospital Carl Gustav Carus, Technische Universität Dresden, Helmholtz-Zentrum Dresden - Rossendorf, Dresden, Germany}
\address[hzdr-ior]{Helmholtz-Zentrum Dresden-Rossendorf, Institute of Radiooncology - OncoRay, Händelallee 26, 01309 Dresden, Germany}
\address[dktk]{German Cancer Consortium DKTK, Partner Site Dresden, Dresden, Germany}
\address[wol-uni]{Centre for Medical Radiation Physics, University of Wollongong, Wolllongong NSW 2522,
Australia}
\address[wol-hos]{Illawarra Cancer Care Centre, Wollongong Hospital, Wollongong NSW 2500, Australia}
\address[hzdr-iorp]{Helmholtz-Zentrum Dresden-Rossendorf, Institute of Radiation Physics, Dresden, Germany}
\address[str]{Department of Radiation Oncology, Faculty of Medicine and University Hospital Carl Gustav Carus, Technische Universität Dresden}

\cortext[cor1]{Corresponding author}
 
\begin{abstract}
Given the sensitivity of proton therapy to anatomical variations, this cancer treatment modality is expected to benefit greatly from integration with  magnetic resonance (MR) imaging. One of the obstacles hindering such an integration are strong magnetic field induced dose distortions. These have been predicted in simulation studies, but no experimental validation has been performed so far. Here we show the first measurement of planar distributions of dose deposited by therapeutic proton pencil beams traversing a one-Tesla transversal magnetic field while depositing energy in a tissue-like phantom using film dosimetry. The lateral beam deflection ranges from one millimeter to one centimeter for 80 to \SI{180} MeV beams. Simulated and measured deflection agree within one millimeter for all studied energies. These results proof that the magnetic field induced proton beam deflection is both measurable and accurately predictable. This demonstrates the feasibility of accurate dose measurement and hence validates dose predictions for the framework of MR-integrated proton therapy. 
\end{abstract}
  
\end{frontmatter}
	
With about half of all cancer patients undergoing radiation therapy, this modality has become one of the three pillars of modern cancer treatment next to chemotherapy and surgery \cite{Baumann2016}. Radiation therapy makes use of ionising radiation, usually high-energetic photons, in order to irreversibly damage the DNA and chromatin of clonogenic tumour cells, and thereby prevent these cells from proliferating. Over the last two decades, the use of accelerated protons has emerged as an important treatment option, especially for tumours of the central nervous system and for paediatric cases \cite{Gondi2016,Durante2017}. As compared to conventional photon-based radiation therapy, proton therapy utilizes the electromagnetic properties of charged particles giving rise to a depth-dose distribution that is characterized by a steep dose maximum at the so-called Bragg peak, with almost no dose beyond the peak. The Bragg peak can be positioned to cover the tumour volume in depth by adjustment of its energy, and laterally by use of dipole magnets or collimation devices. This allows to deliver dose distributions with improved conformality to the tumour volume and hence a better sparing of healthy tissue, especially beyond the tumour.

At the same time, the steepness of the Bragg peak and its positional dependence on the material composition in the beam path make proton treatment far more sensitive to morphological changes in the patient than photon (or X-ray) therapy. Possible anatomical alterations and differences in patient positioning between irradiations (interfractional uncertainties), as well as organ motion and deformation during irradiation (intrafractional
uncertainties, e.g. due to respiration, muscle tension and digestion) lead to considerable uncertainties in the dose delivery. These uncertainties currently translate into large treatment margins around the tumour und thus compromise the dosimetric benefit of proton therapy, which may lead to undesired side effects for the patients (treatment toxicity). 

One strategy to reduce such side effects is to monitor the patient's anatomy during dose delivery in real-time and to adapt the beam accordingly. Providing unmatched soft-tissue contrast, subsecond temporal resolution and freedom from ionizing radiation dose, Magnetic Resonance Imaging (MRI) appears to be an ideal candidate for anatomy monitoring during treatment delivery \cite{Lauterbur1973,Lagendijk2014}. For photon beam therapy, the integration of both modalities has already been realized, and first-generation hybrid systems have recently found their way into clinical application \cite{Mutic2014,Fallone2014,Lagendijk2014a,Keall2014}. 

Because of the high anatomical sensitivity of proton therapy mentioned above, the integration of MRI into proton therapy (MRiPT) is expected to be even more beneficial than for photon therapy and thus has gained interest in the last years \cite{Oborn2017}. However, MRiPT has only been studied on a conceptual level thus far, as a number of technical and physical challenges need to be overcome. Complex mutual interactions are expected between the transient electromagnetic fields of the MR scanner and the proton beam line \cite{Oborn2016}. Furthermore, unlike photons, protons being charged particles are subject to the Lorentz force when traversing the magnetic field of an MRI scanner, causing the beam to be deflected from its otherwise straight path. This is a well-understood and benchmarked phenomenon in vacuum. However, for protons interacting with substances of the human body (e.g. tissue, bone, air, fat, blood), a continuous energy loss will affect the local curvature of the beam as it penetrates the body. So far, only simulation studies exist investigating this effect using particle tracking algorithms in a water phantom or patient geometry. Severe distortions of the dose distribution in the patient have been predicted; particularly, the Bragg peak is expected to be displaced by a few millimeters up to several centimeters, depending on beam energy and magnetic flux density \cite{Raaymakers2008,Wolf2012,Moteabbed2014,Oborn2015,Hartman2015,Fuchs2017,Schellhammer2017}. 

For the design of dose distributions during treatment planning, the dose simulation models predicting these effects need to be validated with measurements. Hence experimental benchmark data is required. However, such data does not exist, thereby inhibiting the employment of the models used in simulation studies for potential clinical use.

For this purpose, measurements are needed involving a clinical proton pencil beam (beam energy \(E_0\) between \SI{70}{MeV} and \SI{230}{MeV}), a magnet whose flux density is in the clinical range (\(B_0 = 0.35...\SI{3}{T})\), a tissue-like phantom (density \(\rho \approx \SI{1}{g cm^{-3}}\)), and a dosimeter having an adequate spatial resolution (\(\Delta x < \SI{1}{mm}\)) as well as a magnetic field-independent dose response. 

Here we describe the first measurement of magnetic field induced dose distortion of a slowing-down therapeutic proton pencil beam within a tissue-like medium, and a comparison to state-of-the-art dose simulations based on Monte Carlo particle tracking.
    
\section*{Proton dose measurement within a magnetic field} \label{sec:main}
\begin{figure*}%
    \centering
    \subfloat[]{{\includegraphics[width=.61\textwidth]{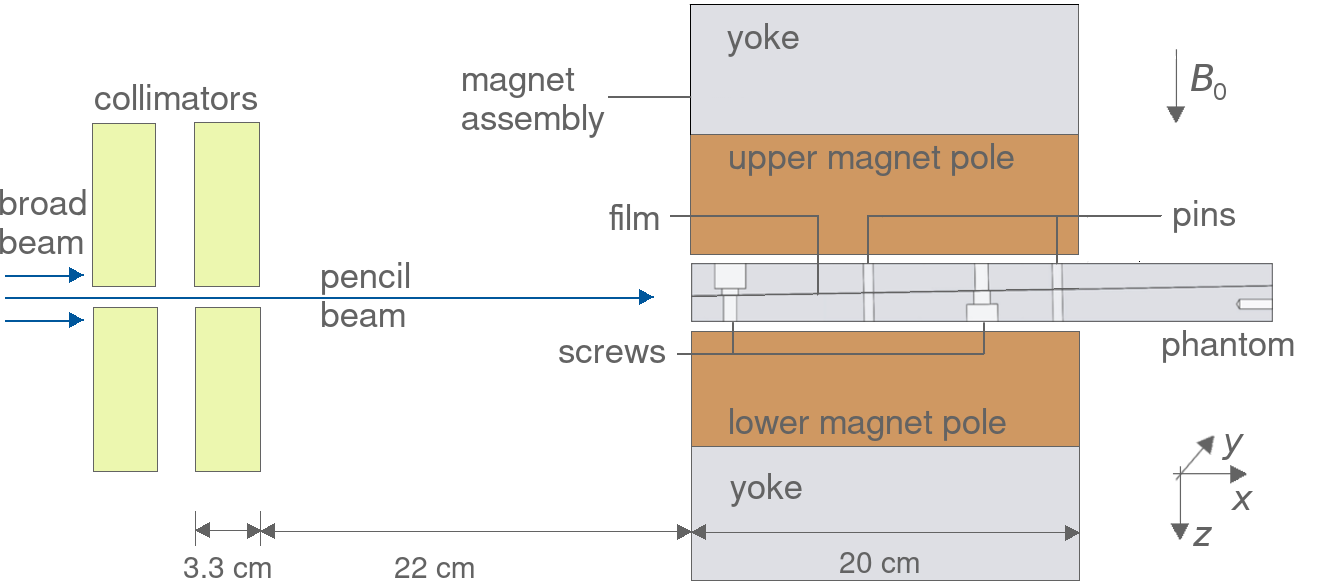}\label{fig:setup_scheme} }}
    \subfloat[]{{\includegraphics[width=.23\textwidth]{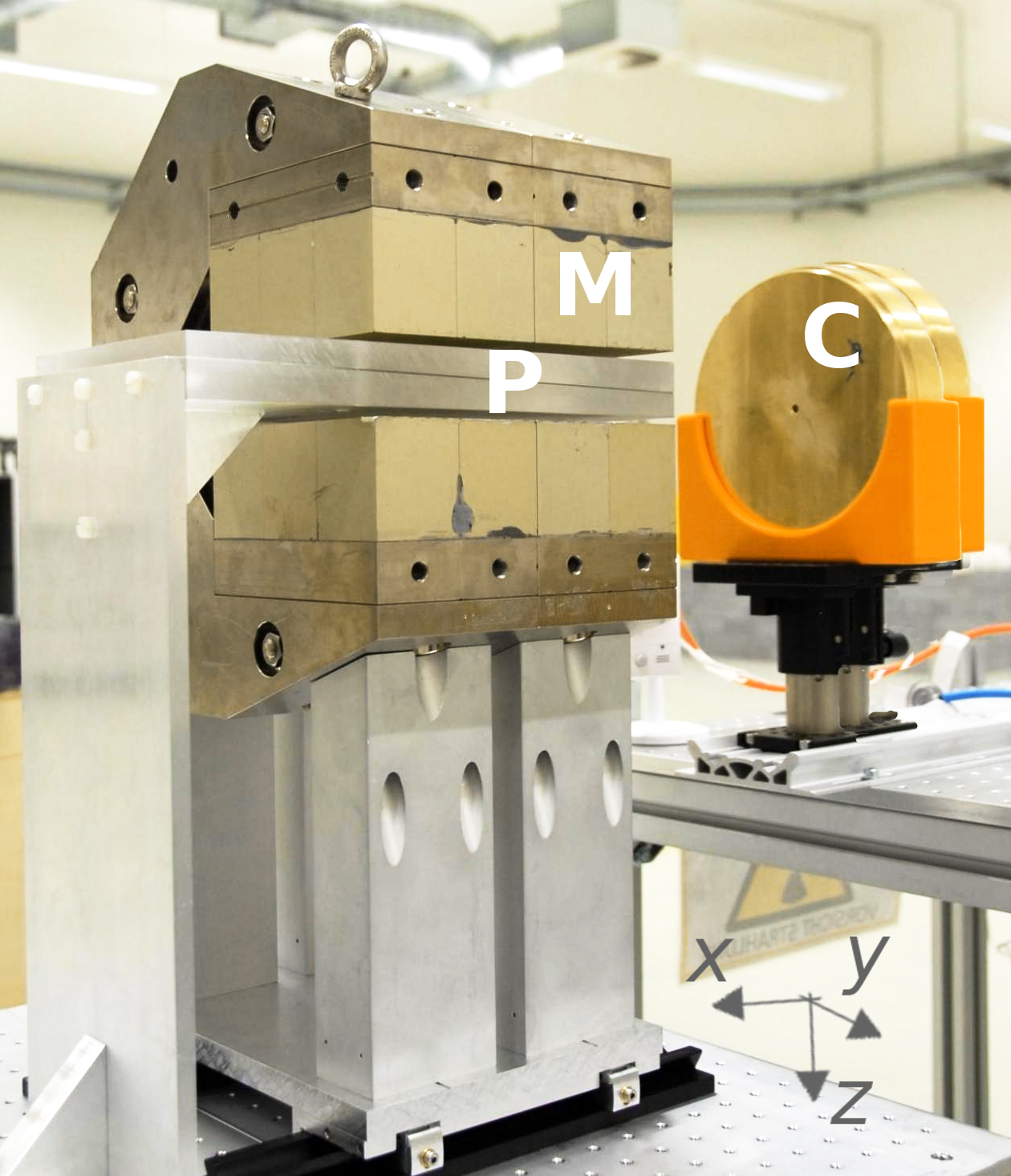}\label{fig:setup_photo} }}%
    \caption{\label{fig:setup} Schematic sagittal view (a) and photograph (b) of the experimental setup. The proton beam passed through collimators (C) and a slab phantom (P) placed inside the field of a permanent magnet (M). The dose in the phantom was measured using a Gafchromic film detector. A left-handed coordinate system is used. Photo edited for clarity.}
\end{figure*}

The measurement setup is depicted in Figure \ref{fig:setup}. It consists of a collimated proton beam,  a slab phantom containing a horizontally placed film dosimeter, and a permanent magnet assembly.

The proton beam was generated by an isochronous cyclotron (C230, IBA, Louvain-La-Neuve, Belgium) at the proton therapy facility of our institution. The horizontal static beam line was used and defined the \(x\)-axis of the setup. Beams of energy \(E_0\) between \SI{80}{MeV} and \SI{180}{MeV} were used and collimated to pencil beams of \SI{1}{cm} diameter in order to prevent the magnet from being damaged from direct radiation exposure.

A slab phantom was produced from polymethyl methacrylate (PMMA, \(\rho = \SI{1.186}{g cm^{-3}}\)) and mounted horizontally in the air gap between the poles of a C-shaped permanent magnet assembly. It consisted of two slabs, such that a film dosimeter was placed horizontally in the central plane parallel to the beam. The contact area of the slabs was bevelled by \(\alpha = \SI{1}{\degree}\) to reduce the dependence of the dose distribution on the film material and a possible air gap between the phantom and film \cite{Zhao2010}. Two vertically oriented pins in the phantom and corresponding holes in the films ensured a reproducible alignment of the film relative to the phantom.

A self-developing Gafchromic EBT3 film detector of \SI{280}{\micro\meter} thickness (Ashland, Covington, USA) was used to measure the planar dose distributions in the central plane of the proton beam, as it provides a submillimeter spatial resolution and is largely unaffected by magnetic fields \cite{Wang2016}. 

The phantom and film detector were placed inside a transversal magnetic field (see Supporting Figure \ref{fig:field}) produced by a magnet assembly consisting of two \chem{Nd_2Fe_{14}B} permanent magnet poles and a yoke. The maximum magnetic flux density of \(B_0 =\SI{0.95}{T}\) was chosen to be comparable to that of existing MR-integrated photon therapy systems with flux densities between \SI{0.35}{T} and \SI{1.5}{T} \cite{Mutic2014,Fallone2014,Lagendijk2014a,Keall2014}. The main field component was aligned parallel to the \(z\)-axis (pointing downwards), and caused a deflection of the proton beam in the (positive) \(y\)-direction.

Prior to the first irradiation experiment, all three magnetic field components were mapped at \SI{5}{mm} resolution using an automated magnetometry setup, comprising of a robotic positioning device and a Hall probe (MMTB-6J04-VG, Lake Shore Cryotronics, Westerville, USA) that was connected to a digital Gaussmeter (Model 421, Lake Shore Cryotronics) \cite{Gantz2017}. To monitor the main field component's temporal stability, point measurements along the central \(x\)-axis at 5, 10 and \SI{15}{cm} depth relative to the phantom's front face were performed after each irradiation experiment. Deviations were within \SI{3}{mT}, which was smaller than the measurement precision (4 \SI{}{mT}), and not systematic.

Further details of the experimental setup can be found in Supporting \ref{sec:expsetup}.

\begin{figure}[t]
\begin{center}
 \subfloat[]{\includegraphics[width=0.5\textwidth]{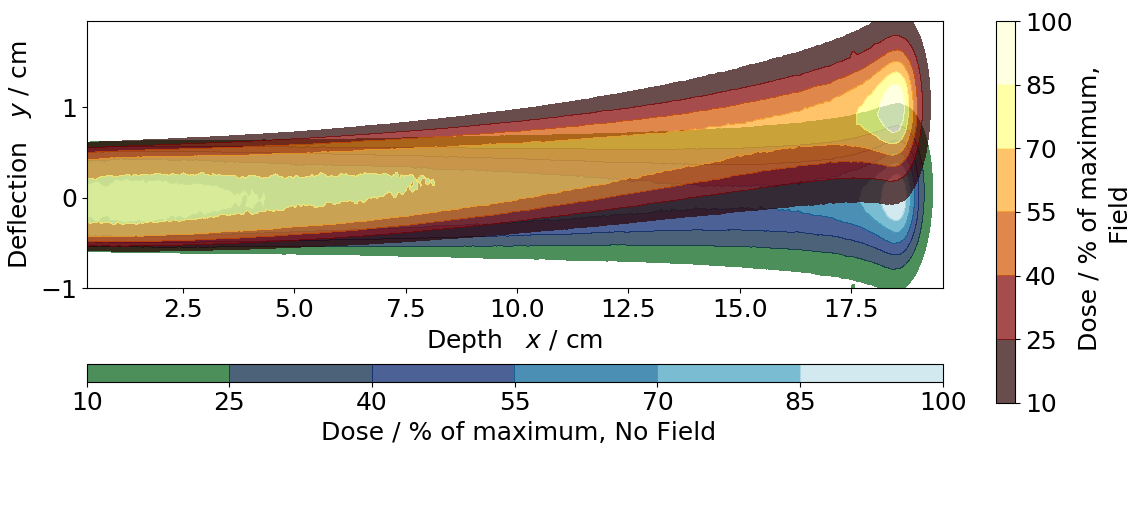}}
 \subfloat[]{\includegraphics[width=0.5\textwidth]{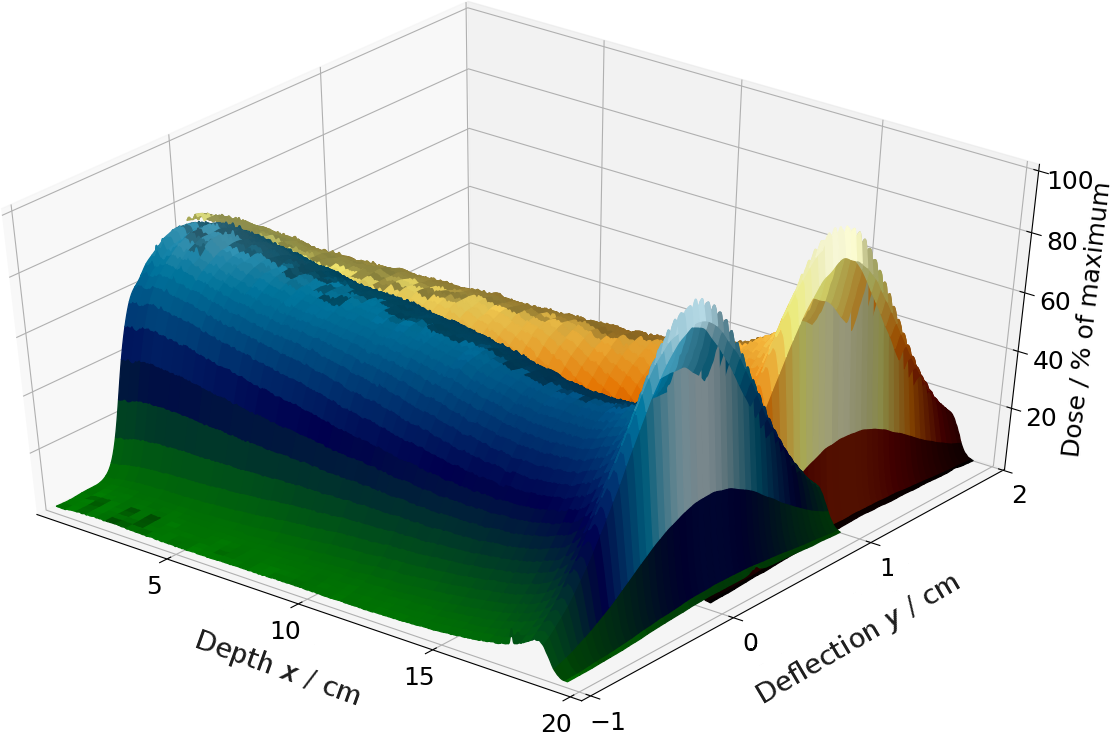}}
\caption{Relative dose distribution of a 180 MeV proton beam in PMMA on the film dosimeter, without (redscale) and with (bluescale) magnetic field. (a): Top view, (b): side view. \label{fig:Comparison3D}}
\label{fig:3D}
\end{center}
\end{figure}

\begin{figure}[ht]
\begin{center}
\hspace{4mm}\includegraphics[width=0.695\textwidth]{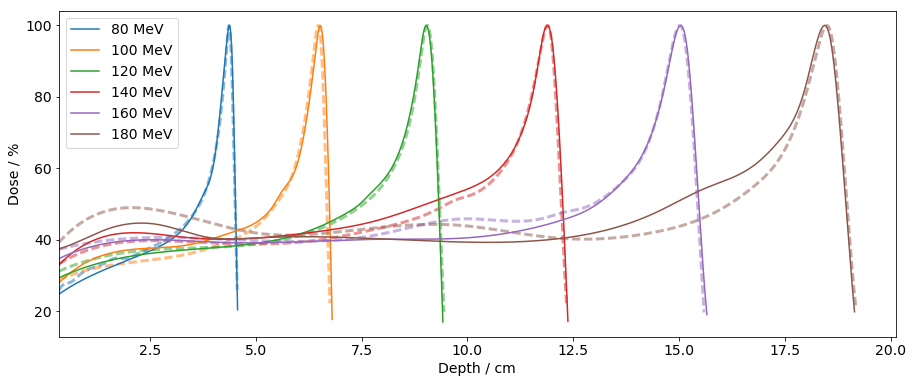}
 \caption{Relative measured depth-dose curves with (solid lines) and without (dashed) magnetic field.}
 \label{fig:Depthdose}
\end{center}
\end{figure}

\begin{figure}[b]
\begin{center}
\includegraphics[width=0.55\textwidth]{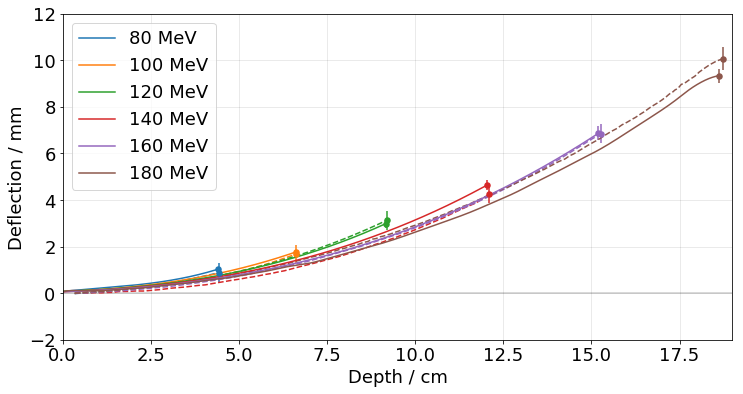}
 \caption{Predicted (solid lines) and measured (dashed) deflected central beam trajectories for different beam energies. Error bars refer to systematic and statistical uncertainties given in Supplement Tables \ref{tab:expuncertainty} and \ref{tab:simuncertainty}.}
\label{fig:trajectories}
\end{center}
\end{figure}

\section*{Beam deflection and Bragg peak dislocation} \label{cha:results}

The influence of the magnetic field on the proton dose distribution could be observed on the film dosimeter signal, as depicted for \SI{180}{MeV} in Figure \ref{fig:Comparison3D}. The characteristic shape of the proton dose distribution was visible both with and without magnetic field, but with magnetic field the beam followed a deflected trajectory resulting in a laterally shifted Bragg peak position.

To investigate the impact of the magnetic field on the range of the beam (i.e. Bragg peak retraction), depth-dose curves were reconstructed from the film dose distributions by a radial integration method (see Supporting \ref{sec:integration} and Figure \ref{fig:Depthdose}). From these, the proton range \(R_{80}\), defined as the penetration depth of the beam at the \SI{80}{\%} distal end of the Bragg peak maximum, was extracted both with and without magnetic field. Bragg peak retraction was found to be below \SI{0.5}{mm} for all studied energies. Small deviations were found between depth-dose curves with and without magnetic field in the plateau region (see Figure \ref{fig:Depthdose}), however, repeated experiments showed that these were due to statistical setup uncertainties (see Supporting \ref{sec:simuncertainty}) rather than the magnetic field. No magnetic-field induced change in the absolute measured dose was observed, showing the suitability of Gafchromic EBT3 film dosimeters for this experiment. 

To quantify the lateral Bragg peak dislocation, the central trajectory of the deflected beam was reconstructed using a depth layer-wise Gaussian fitting method (see Supporting \ref{sec:integration} and Figure \ref{fig:trajectories}). The extracted Bragg peak dislocation ranged from \SI{1}{mm} for \SI{80}{MeV} up to \SI{10}{mm} for \SI{180}{MeV}. 

Note that the deflected beam trajectories did not coincide for different energies, but fanned out energy-dependently, as the radius of beam path increases with increasing proton energy \cite{Wolf2012,Schellhammer2017}. For high energies (most prominently \SI{180}{MeV}), trajectories appeared inflected in the opposite direction towards the end of the beam path. This can be explained by a combination of two effects. Firstly, the radial integration planes are increasingly less perpendicular to the beam direction. Secondly, within the energy spectrum of the beam, protons with higher energies are less deflected by the Lorentz force than those with smaller energies. Thus, at high depths, only long range (high initial energy) protons remain, which shift the central beam trajectory towards smaller deflection (energy separation effect \cite{Fuchs2017}). 

Measured deflection and retraction values and their uncertainties are detailed in Supporting Table \ref{tab:deflection}.

\section*{Agreement with computational model} \label{sec:agreement}

\begin{figure}%
    \centering
    \subfloat[]{{\includegraphics[width=.49\textwidth]{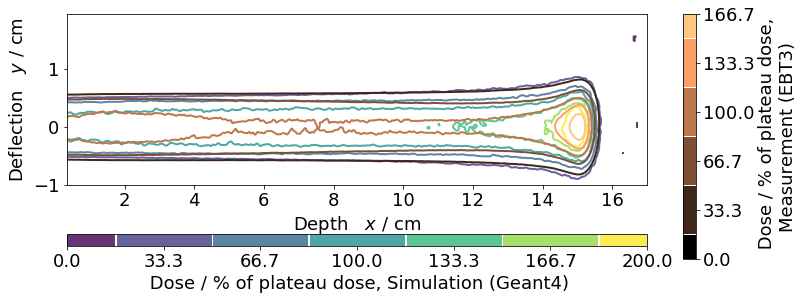}}}
    \hspace{2mm}
    \subfloat[]{{\includegraphics[width=.49\textwidth]{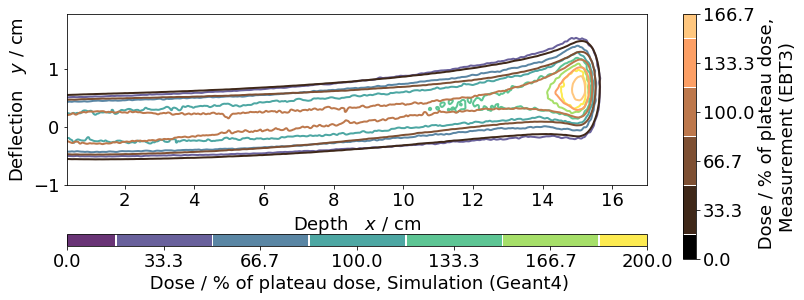}}}
    \caption{\label{fig:2D} Top panel: Predicted (green) and measured (copper) dose distribution of a \SI{160}{MeV} proton beam in PMMA without (a) and with magnetic field (b). Dose is given in percent relative to the median dose in the central plateau region (\( \SI{1.9}{cm}\leq x\leq\SI{2.1}{cm}\) and \( \SI{-0.5}{cm}\leq y\leq\SI{0.5}{cm}\)).}
\end{figure}

The measured data allowed for a verification of the dose prediction in magnetic fields by Monte Carlo based particle tracking simulations, being considered as the gold standard for dose prediction in radiation therapy  planning. For this study, such simulations were performed using the Geant4 toolkit version 10.2.p02 \cite{Agostinelli2003,Allison2006}. Details on the simulation beam model can be found in Supporting \ref{sec:beammodel}.

The shape of the predicted dose distributions agreed with the measured distributions, as shown exemplarily for a \SI{160}{MeV} proton beam in Figure \ref{fig:2D}. Due to saturation of the film signal caused by the particles' increasing energy deposition per path length, differences between predicted and measured dose increased towards the Bragg peak. The difference in the absolute Bragg peak dose was consistent with the findings of previous studies \cite{Zhao2010,Arjomandy2012,Perles2013}, and was not increased by the magnetic field. Thus, the introduction of a magnetic field did not compromise the dose prediction accuracy.

\begin{figure}
\begin{center}

 \includegraphics[width=.55\textwidth]{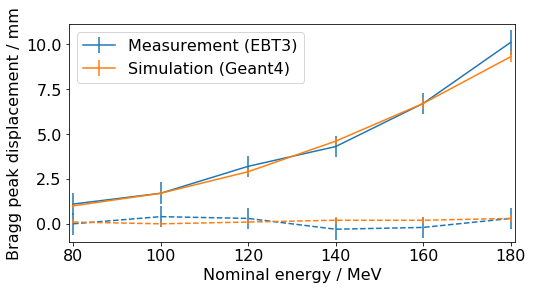}
 \caption{Predicted and measured Bragg peak displacement in lateral (solid) and longitudinal (dashed) direction for discrete beam energy levels. Error bars refer to systematic and statistical uncertainties given in Supplement Tables \ref{tab:expuncertainty} and \ref{tab:simuncertainty}. Individual data points are interconnected for visualization only.}
\label{fig:BPposistionComparison}
\end{center}
\end{figure}

To quantify the accuracy of the calculated Bragg peak position, its retraction and deflection were compared (see Figure \ref{fig:BPposistionComparison}). Both quantities agreed within measurement and simulation uncertainties (\SI{0.8}{mm}) for all studied energies. There was no systematic difference in Bragg peak retraction and deflection between the simulation and measurement data. 

\section*{Conclusions}

For the first time, magnetic field induced dose distortions of a slowing-down therapeutic proton beam in a tissue-like medium have been measured and compared against particle tracking simulations. In a transverse magnetic field of \SI{0.95}{T}, it was shown that lateral Bragg peak displacement ranges between \SI{1}{mm} and \SI{10}{mm} for proton energies between 80 and \SI{180}{MeV} in PMMA. Range retraction was found to be \(\leq \SI{0.5}{mm}\), which is clinically acceptable. These results were shown to agree within \SI{0.8}{mm} with Monte Carlo based particle tracking simulations.

The obtained range of \SI{1}{mm} to \SI{10}{mm} Bragg peak displacement corresponds very well with previous theoretical studies in a homogeneous magnetic field in water \cite{Fuchs2017,Schellhammer2017}. For inhomogeneous media, a magnetic field induced dose increase at medium-air material boundaries in the order of \SI{2}{\%} has been predicted caused by electrons returning to the material boundary after being deflected by Lorentz force in air (electron return effect \cite{Fuchs2017}). The experimental verification thereof is subject of an ongoing study.  

This study showed for the first time that the magnetic field induced proton dose distortion is both measurable and accurately predictable in a tissue-like beam stopping medium. This validates the models used by Monte Carlo simulations and allows them to be used for accurate dose prediction in future treatment planning studies on MR-integrated proton therapy.
    
  \section*{Methods}
Details on the experimental setup and the Monte Carlo model as well as an estimation of sources of uncertainty can be found in the Supplementary Information section. 

  \section*{Ackowledgements}
The authors thank Wolfgang Enghardt, Stephan Helmbrecht, Jörg Pawelke, Elke Beyreuther and Leonard Karsch (OncoRay, Germany) for discussing the experimental setup. We thank Patrick Wohlfahrt (OncoRay, Germany) for providing depth-dose curves of the beam line, Florian Brack (HZDR, Dresden) for providing the magnetic finite element model, and Susanna Guatelli (UOW, Australia), Henrik Schulz, Richard Pausch, Carlchristian Eckert and Alexander Matthes (HZDR, Germany) for support with the Geant4 software and the high performance cluster. We thank Jan Pipek (INFN, Czech Republic) for testing and implementing our bug fix to the next Geant4 release, and Susanna Guatelli (UOW, Australia), Andreas Resch (MedUni Wien, Austria), Brent Huisman (INSA, France) and Loic Grevillot (MedAustron, Austria) for advice on Geant4 physics lists. We thank Michael Knobel (HZDR, Dresden) for performing the phantom density determination, and Norbert M\"ullner (Vacuumschmelze GmbH, Germany) for providing data on the magnet material composition. We thank Karl Zeil (HZDR, Dresden, Germany) for providing the magnet assembly, Robert Schönert (HZDR, Dresden, Germany) for manufacturing the phantom and Elke Beyreuther (HZDR, Dresden, Germany) for practical advice on film dosimetry. 
    
  \section*{Author contributions}
S.M.S. and A.L.H. conceptually designed the study. S.M.S., S.G., A.L. and A.L.H. planned and optimized the experiments; S.M.S, B.O., A.L., and M.B. planned and optimized the simulations. S.M.S., S.G. and A.L. conducted the experiments; S.M.S. conducted the simulations. S.M.S and S.G. performed the data analysis. S.M.S. performed the comparison between experiment and simulation. All authors contributed to data interpretation. S.M.S. drafted the manuscript of this article. All authors critically revised the article and approved the final version submitted for publication.

\section*{Competing Financial Interest Statement}
The authors declare no competing financial interests.

  \section*{References}
  
   \bibliographystyle{naturemag}
   \bibliography{references}

\begin{thebibliography}{10}
\expandafter\ifx\csname url\endcsname\relax
  \def\url#1{\texttt{#1}}\fi
\expandafter\ifx\csname urlprefix\endcsname\relax\def\urlprefix{URL }\fi
\providecommand{\bibinfo}[2]{#2}
\providecommand{\eprint}[2][]{\url{#2}}

\bibitem{Baumann2016}
\bibinfo{author}{Baumann, M.} \emph{et~al.}
\newblock \bibinfo{title}{Radiation oncology in the era of precision medicine}.
\newblock \emph{\bibinfo{journal}{Nat. Rev. Cancer}}
  \textbf{\bibinfo{volume}{16}}, \bibinfo{pages}{234--249}
  (\bibinfo{year}{2016}).

\bibitem{Gondi2016}
\bibinfo{author}{Gondi, V.}, \bibinfo{author}{Yock, T.~I.} \&
  \bibinfo{author}{Mehta, M.~P.}
\newblock \bibinfo{title}{Proton therapy for paediatric {CNS} tumours -
  improving treatment-related outcomes}.
\newblock \emph{\bibinfo{journal}{Nat. Rev. Neurol.}}
  \textbf{\bibinfo{volume}{12}}, \bibinfo{pages}{334--345}
  (\bibinfo{year}{2016}).

\bibitem{Durante2017}
\bibinfo{author}{Durante, M.}, \bibinfo{author}{Orecchia, R.} \&
  \bibinfo{author}{Loeffler, J.~S.}
\newblock \bibinfo{title}{Charged-particle therapy in cancer: clinical uses and
  future perspectives}.
\newblock \emph{\bibinfo{journal}{Nat. Rev. Clin. Oncol.}}
  \textbf{\bibinfo{volume}{14}}, \bibinfo{pages}{483--495}
  (\bibinfo{year}{2017}).

\bibitem{Lauterbur1973}
\bibinfo{author}{Lauterbur, P.~C.}
\newblock \bibinfo{title}{Image formation by induced local interactions:
  examples employing nuclear magnetic resonance}.
\newblock \emph{\bibinfo{journal}{Nature}} \textbf{\bibinfo{volume}{242}},
  \bibinfo{pages}{190--191} (\bibinfo{year}{1973}).

\bibitem{Lagendijk2014}
\bibinfo{author}{Lagendijk, J. J.~W.} \emph{et~al.}
\newblock \bibinfo{title}{{MR guidance in radiotherapy}}.
\newblock \emph{\bibinfo{journal}{Phys. Med. Biol.}}
  \textbf{\bibinfo{volume}{59}}, \bibinfo{pages}{R349--R369}
  (\bibinfo{year}{2014}).

\bibitem{Mutic2014}
\bibinfo{author}{Mutic, S.} \& \bibinfo{author}{Dempsey, J.}
\newblock \bibinfo{title}{The {ViewRay} system: Magnetic resonance-guided and
  controlled radiotherapy}.
\newblock \emph{\bibinfo{journal}{Sem. Rad. Onc.}}  (\bibinfo{year}{2014}).

\bibitem{Fallone2014}
\bibinfo{author}{Fallone, B.~G.}
\newblock \bibinfo{title}{{The Rotating Biplanar Linac-Magnetic Resonance
  Imaging System}}.
\newblock \emph{\bibinfo{journal}{Semin. Radiat. Oncol.}}
  \textbf{\bibinfo{volume}{24}}, \bibinfo{pages}{200--202}
  (\bibinfo{year}{2014}).

\bibitem{Lagendijk2014a}
\bibinfo{author}{Lagendijk, J.~J.}, \bibinfo{author}{Raaymakers, B.~W.} \&
  \bibinfo{author}{van Vulpen, M.}
\newblock \bibinfo{title}{{The magnetic resonance imaging-linac system}}.
\newblock \emph{\bibinfo{journal}{Semin. Radiat. Oncol.}}
  \textbf{\bibinfo{volume}{24}}, \bibinfo{pages}{207--9}
  (\bibinfo{year}{2014}).

\bibitem{Keall2014}
\bibinfo{author}{Keall, P.~J.}, \bibinfo{author}{Barton, M.} \&
  \bibinfo{author}{Crozier, S.}
\newblock \bibinfo{title}{{The Australian Magnetic Resonance Imaging-Linac
  Program}}.
\newblock \emph{\bibinfo{journal}{Semin. Radiat. Oncol.}}
  \textbf{\bibinfo{volume}{24}}, \bibinfo{pages}{203--206}
  (\bibinfo{year}{2014}).

\bibitem{Oborn2017}
\bibinfo{author}{Oborn, B.~M.} \emph{et~al.}
\newblock \bibinfo{title}{{Future of medical physics: real-time MRI guided
  proton therapy}}.
\newblock \emph{\bibinfo{journal}{Med. Phys.}} \textbf{\bibinfo{volume}{44}},
  \bibinfo{pages}{e77--e90} (\bibinfo{year}{2017}).

\bibitem{Oborn2016}
\bibinfo{author}{Oborn, B.~M.} \emph{et~al.}
\newblock \emph{\bibinfo{title}{{MRI Guided Proton Therapy: Pencil beam
  scanning in an MRI fringe field}}}.
\newblock \bibinfo{organization}{ICTR-PHE}, \bibinfo{address}{Geneva, February
  15-19} (\bibinfo{year}{2016}).

\bibitem{Raaymakers2008}
\bibinfo{author}{Raaymakers, B.}, \bibinfo{author}{Raaijmakers, A. J.~E.} \&
  \bibinfo{author}{Lagendijk, J.~J.}
\newblock \bibinfo{title}{{Feasibility of MRI guided proton therapy: magnetic
  field dose effects}}.
\newblock \emph{\bibinfo{journal}{Phys. Med. Biol.}}
  \textbf{\bibinfo{volume}{53}}, \bibinfo{pages}{5615--5622}
  (\bibinfo{year}{2008}).

\bibitem{Wolf2012}
\bibinfo{author}{Wolf, R.} \& \bibinfo{author}{Bortfeld, T.}
\newblock \bibinfo{title}{{An analytical solution to proton Bragg peak
  deflection in a magnetic field}}.
\newblock \emph{\bibinfo{journal}{Phys. Med. Biol.}}
  \textbf{\bibinfo{volume}{57}}, \bibinfo{pages}{N329--N337}
  (\bibinfo{year}{2012}).

\bibitem{Moteabbed2014}
\bibinfo{author}{Moteabbed, M.}, \bibinfo{author}{Schuemann, J.} \&
  \bibinfo{author}{Paganetti, H.}
\newblock \bibinfo{title}{{Dosimetric feasibility of real-time MRI-guided
  proton therapy}}.
\newblock \emph{\bibinfo{journal}{Med. Phys.}} \textbf{\bibinfo{volume}{41}},
  \bibinfo{pages}{111713 1--11} (\bibinfo{year}{2014}).

\bibitem{Oborn2015}
\bibinfo{author}{Oborn, B.~M.} \emph{et~al.}
\newblock \bibinfo{title}{{Proton beam deflection in MRI fields: Implications
  for MRI-guided proton therapy}}.
\newblock \emph{\bibinfo{journal}{Med. Phys.}} \textbf{\bibinfo{volume}{42}},
  \bibinfo{pages}{2113--2124} (\bibinfo{year}{2015}).

\bibitem{Hartman2015}
\bibinfo{author}{Hartman, J.} \emph{et~al.}
\newblock \bibinfo{title}{{Dosimetric feasibility of intensity modulated proton
  therapy in a transverse magnetic field of 1.5 T}}.
\newblock \emph{\bibinfo{journal}{Phys. Med. Biol.}}
  \textbf{\bibinfo{volume}{60}}, \bibinfo{pages}{5955--5969}
  (\bibinfo{year}{2015}).

\bibitem{Fuchs2017}
\bibinfo{author}{Fuchs, H.}, \bibinfo{author}{Moser, P.},
  \bibinfo{author}{Gröschl, M.} \& \bibinfo{author}{Georg, D.}
\newblock \bibinfo{title}{{Magnetic field effects on particle beams and their
  implications for dose calculation in MR guided particle therapy}}.
\newblock \emph{\bibinfo{journal}{Medical Physics}}
  \textbf{\bibinfo{volume}{44}}, \bibinfo{pages}{1149--1156}
  (\bibinfo{year}{2017}).

\bibitem{Schellhammer2017}
\bibinfo{author}{Schellhammer, S.~M.} \& \bibinfo{author}{Hoffmann, A.~L.}
\newblock \bibinfo{title}{{Prediction and compensation of magnetic beam
  deflection in MR-integrated proton therapy: A method optimized regarding
  accuracy, versatility and speed}}.
\newblock \emph{\bibinfo{journal}{Phys. Med. Biol.}}
  \textbf{\bibinfo{volume}{62}}, \bibinfo{pages}{1548--1564}
  (\bibinfo{year}{2017}).

\bibitem{Zhao2010}
\bibinfo{author}{Zhao, L.} \& \bibinfo{author}{Das, I.~J.}
\newblock \bibinfo{title}{{Gafchromic EBT film dosimetry in proton beams}}.
\newblock \emph{\bibinfo{journal}{Phys. Med. Biol.}}
  \textbf{\bibinfo{volume}{55}}, \bibinfo{pages}{N291--N301}
  (\bibinfo{year}{2010}).

\bibitem{Wang2016}
\bibinfo{author}{Wang, J.}, \bibinfo{author}{Rubinstein, A.},
  \bibinfo{author}{Ohrt, J.}, \bibinfo{author}{Ibbott, G.} \&
  \bibinfo{author}{Wen, Z.}
\newblock \bibinfo{title}{{Effect of a strong magnetic field on TLDs, OSLDs,
  and Gafchromic films using an electromagnet}}.
\newblock \emph{\bibinfo{journal}{Med. Phys.}} \textbf{\bibinfo{volume}{43}},
  \bibinfo{pages}{3873--3874} (\bibinfo{year}{2016}).

\bibitem{Gantz2017}
\bibinfo{author}{Gantz, S.}
\newblock \emph{\bibinfo{title}{{Characterization of the magnetic field of a
  0.95 T permanent dipole magnet and experimental validation of proton beam
  deflection in a magnetic field within tissue-equivalent material}}}.
\newblock Master's thesis, \bibinfo{school}{TU Dresden} (\bibinfo{year}{2017}).

\bibitem{Agostinelli2003}
\bibinfo{author}{Agostinelli, S.} \emph{et~al.}
\newblock \bibinfo{title}{{Geant4 - a simulation toolkit}}.
\newblock \emph{\bibinfo{journal}{Nuclear Instruments and Methods in Physics
  Research Section A}} \textbf{\bibinfo{volume}{506}}, \bibinfo{pages}{250 --
  303} (\bibinfo{year}{2003}).

\bibitem{Allison2006}
\bibinfo{author}{Allison, J.} \emph{et~al.}
\newblock \bibinfo{title}{{Geant4 developments and applications}}.
\newblock \emph{\bibinfo{journal}{IEEE Transactions on Nuclear Science}}
  \textbf{\bibinfo{volume}{53}}, \bibinfo{pages}{270--278}
  (\bibinfo{year}{2006}).

\bibitem{Arjomandy2012}
\bibinfo{author}{Arjomandy, B.}, \bibinfo{author}{Tailor, R.},
  \bibinfo{author}{Zhao, L.} \& \bibinfo{author}{Devic, S.}
\newblock \bibinfo{title}{{EBT2 film as a depth-dose measurement tool for
  radiotherapy beams over a wide range of energies and modalities}}.
\newblock \emph{\bibinfo{journal}{Med. Phys.}} \textbf{\bibinfo{volume}{39}},
  \bibinfo{pages}{912--921} (\bibinfo{year}{2012}).

\bibitem{Perles2013}
\bibinfo{author}{Perles, L.~A.}, \bibinfo{author}{Mirkovic, D.},
  \bibinfo{author}{Anad, A.}, \bibinfo{author}{Titt, U.} \&
  \bibinfo{author}{Mohan, R.}
\newblock \bibinfo{title}{{LET dependence of the response of EBT2 films in
  proton dosimetry modeled as a bimolecular chemical reaction}}.
\newblock \emph{\bibinfo{journal}{Phys. Med. Biol.}}
  \textbf{\bibinfo{volume}{58}}, \bibinfo{pages}{8477--8491}
  (\bibinfo{year}{2013}).

\bibitem{Vacodym2014}
\bibinfo{author}{{Vacuumschmelze}}.
\newblock \emph{\bibinfo{title}{{Selten-Erd-Dauermagnete Vacodym Vacomax}}}.
\newblock \bibinfo{address}{Gr{\"u}ner Weg 37, 63450 Hanau/Germany}
  (\bibinfo{year}{2014}).

\bibitem{Reynoso2016}
\bibinfo{author}{Reynoso, F.~J.} \emph{et~al.}
\newblock \bibinfo{title}{{Technical Note: Magnetic field effects on
  Gafchromic-film response in MR-IGRT}}.
\newblock \emph{\bibinfo{journal}{Med. Phys.}} \textbf{\bibinfo{volume}{43}},
  \bibinfo{pages}{6552--6556} (\bibinfo{year}{2016}).

\bibitem{Zeil2009}
\bibinfo{author}{Zeil, K.}, \bibinfo{author}{Beyreuther, E.},
  \bibinfo{author}{Lessmann, E.}, \bibinfo{author}{Wagner, W.} \&
  \bibinfo{author}{Pawelke, J.}
\newblock \bibinfo{title}{{Cell irradiation setup and dosimetry for
  radiobiological studies at ELBE}}.
\newblock \emph{\bibinfo{journal}{Nuclear Instruments and Methods in Physics
  Research B}} \textbf{\bibinfo{volume}{267}}, \bibinfo{pages}{2403--2410}
  (\bibinfo{year}{2009}).

\bibitem{Sorriaux2012}
\bibinfo{author}{Sorriaux, J.} \emph{et~al.}
\newblock \bibinfo{title}{{Evaluation of Gafchromic EBT3 films characteristics
  in therapy photon, electron and proton beams}}.
\newblock \emph{\bibinfo{journal}{Physica Medica}}  (\bibinfo{year}{2012}).

\bibitem{Wohlfahrt2017}
\bibinfo{author}{Wohlfahrt, P.}, \bibinfo{author}{M{\"o}hler, C.},
  \bibinfo{author}{Richter, C.} \& \bibinfo{author}{Greilich, S.}
\newblock \bibinfo{title}{Evaluation of stopping-power prediction by dual- and
  single-energy computed tomography in an anthropomorphic ground-truth
  phantom}.
\newblock \emph{\bibinfo{journal}{IJROBP}}  (\bibinfo{year}{2017}).
\newblock \bibinfo{note}{Under review}.

\bibitem{Bortfeld1997}
\bibinfo{author}{Bortfeld, T.}
\newblock \bibinfo{title}{{An analytical approximation of the Bragg curve for
  therapeutic proton beams}}.
\newblock \emph{\bibinfo{journal}{Med. Phys.}} \textbf{\bibinfo{volume}{24}},
  \bibinfo{pages}{2024--2033} (\bibinfo{year}{1997}).

\bibitem{PSTAR}
\bibinfo{author}{Berger, M.~J.}, \bibinfo{author}{Coursey, J.~S.},
  \bibinfo{author}{Zucker, M.~A.} \& \bibinfo{author}{Chang, J.}
\newblock \emph{\bibinfo{title}{{ESTAR, PSTAR, and ASTAR: Computer Programs for
  Calculating Stopping-Power and Range Tables for Electrons, Protons, and
  Helium Ions}}}.
\newblock \bibinfo{organization}{National Institute of Standards and
  Technology}, \bibinfo{address}{Gaithersburg, MD} (\bibinfo{year}{2005}).
\newblock \bibinfo{note}{{Version 1.2.3}, Retrieved Apr 27, 2015}.

\bibitem{Almhagen2015}
\bibinfo{author}{Almhagen, E.}
\newblock \emph{\bibinfo{title}{{Development and validation of a scanned proton
  beam model for dose distribution verification using Monte Carlo}}}.
\newblock Master's thesis, \bibinfo{school}{Stockholm University}
  (\bibinfo{year}{2015}).

\bibitem{Geant4Reference}
\bibinfo{author}{{Geant4 collaboration}}.
\newblock \emph{\bibinfo{title}{{Reference Physics Lists}}}
  (\bibinfo{year}{2013}).
\newblock
  \urlprefix\url{http://geant4.cern.ch/support/proc_mod_catalog/physics_lists/referencePL.shtml}.
\newblock \bibinfo{note}{Retrieved Aug 18, 2016}.

\bibitem{Geant4HADRO}
\bibinfo{author}{{Geant4 collaboration}}.
\newblock \emph{\bibinfo{title}{{Geant4 10.1 Release Notes}}}
  (\bibinfo{year}{2014}).
\newblock
  \urlprefix\url{http://geant4.cern.ch/support/ReleaseNotes4.10.1.html}.
\newblock \bibinfo{note}{Retrieved Aug 18, 2016}.

\bibitem{Grevillot2010}
\bibinfo{author}{Grevillot, L.} \emph{et~al.}
\newblock \bibinfo{title}{{Optimization of GEANT4 settings for Proton Pencil
  Beam Scanning simulations using GATE}}.
\newblock \emph{\bibinfo{journal}{Nuclear Instruments and Methods in Physics
  Research Section B}} \textbf{\bibinfo{volume}{268}},
  \bibinfo{pages}{3295--3305} (\bibinfo{year}{2010}).

\bibitem{GATEphysics}
\bibinfo{author}{{GATE collaboration}}.
\newblock \emph{\bibinfo{title}{{Documentation and Recommendations for Users}}}
  (\bibinfo{year}{2017}).
\newblock \urlprefix\url{http://www.opengatecollaboration.org/UsersGuide}.
\newblock \bibinfo{note}{Retrieved Aug 18, 2016}.

\bibitem{Jarlskog2008}
\bibinfo{author}{Jarlskog, C.~Z.} \& \bibinfo{author}{Paganetti, H.}
\newblock \bibinfo{title}{{Physics settings for using the Geant4 toolkit in
  proton therapy}}.
\newblock \emph{\bibinfo{journal}{IEEE Transactions on Nuclear Science}}
  \textbf{\bibinfo{volume}{55}}, \bibinfo{pages}{1018--1025}
  (\bibinfo{year}{2008}).

\bibitem{ICRU73}
\bibinfo{author}{{ICRU}}.
\newblock \bibinfo{title}{{Stopping of ions heavier than Helium}}.
\newblock \emph{\bibinfo{journal}{ICRU Report 73}}  (\bibinfo{year}{2009}).

\bibitem{James1990}
\bibinfo{author}{James, F.}
\newblock \bibinfo{title}{A review of pseudorandom number generators}.
\newblock \emph{\bibinfo{journal}{Comput. Phys. Commun.}}
  \textbf{\bibinfo{volume}{60}}, \bibinfo{pages}{329--344}
  (\bibinfo{year}{1990}).

\bibitem{bug}
\bibinfo{author}{Schellhammer, S.~M.}, \bibinfo{author}{Matthes, A.},
  \bibinfo{author}{Pipek, J.} \& \bibinfo{author}{Apostolakis, J.}
\newblock \emph{\bibinfo{title}{{Problem 1879 - Segfault at magnetic field
  edges}}} (\bibinfo{year}{2017}).
\newblock \urlprefix\url{https://bugzilla-geant4.kek.jp/show_bug.cgi?id=1879}.
\newblock \bibinfo{note}{Retrieved Nov 15, 2016}.

\bibitem{Kurosu2014}
\bibinfo{author}{Kurosu, K.}, \bibinfo{author}{Takashina, M.},
  \bibinfo{author}{Koizumi, M.}, \bibinfo{author}{Das, I.~J.} \&
  \bibinfo{author}{Moskvin, V.}
\newblock \bibinfo{title}{{Optimization of GATE and PHITS Monte Carlo code
  parameters for uniform scanning proton beam based on simulation with FLUKA
  general-purpose code}}.
\newblock \emph{\bibinfo{journal}{Nuclear Instruments and Methods in Physics
  Research Section B}} \textbf{\bibinfo{volume}{336}}, \bibinfo{pages}{45--54}
  (\bibinfo{year}{2014}).

\end{thebibliography}
   References \cite{Vacodym2014} to [42] are referred to in the Supplementary Information section only.

\newpage
  
\newpage
\appendix
\onecolumn
\section*{Supplementary Information}
Details on the material and methods used for this study are given in the following sections.

\section{Experimental measurements}

\subsection{Experimental setup} \label{sec:expsetup}
The experimental setup is displayed in Figure \ref{fig:setup}. Details are given in this section.

The beam defines the \(x\)-axis of the setup. Energies between \SI{80}{} and \SI{180}{MeV} were used to place the Bragg peak inside the main magnetic field (as would be the case in a clinical situation of MR-integrated proton therapy). 

\begin{figure}%
    \centering
    \subfloat[]{{\includegraphics[width=.38\textwidth]{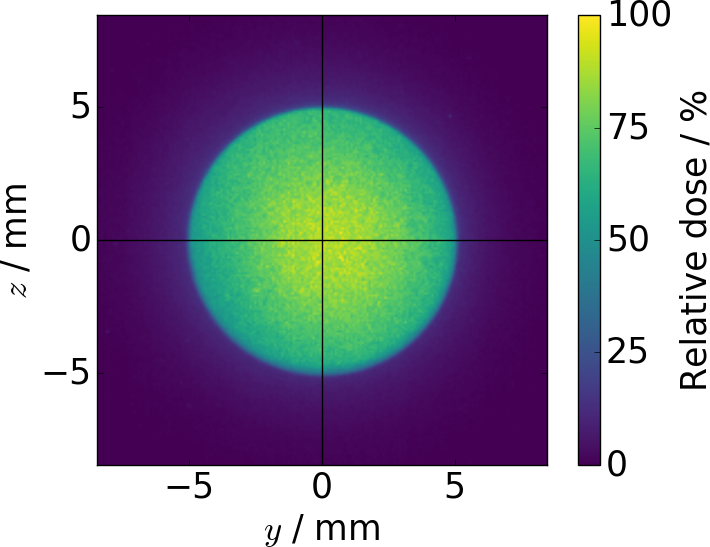} \label{fig:2DBeamProfile} }}    
    \subfloat[]{{\includegraphics[width=.60\textwidth]{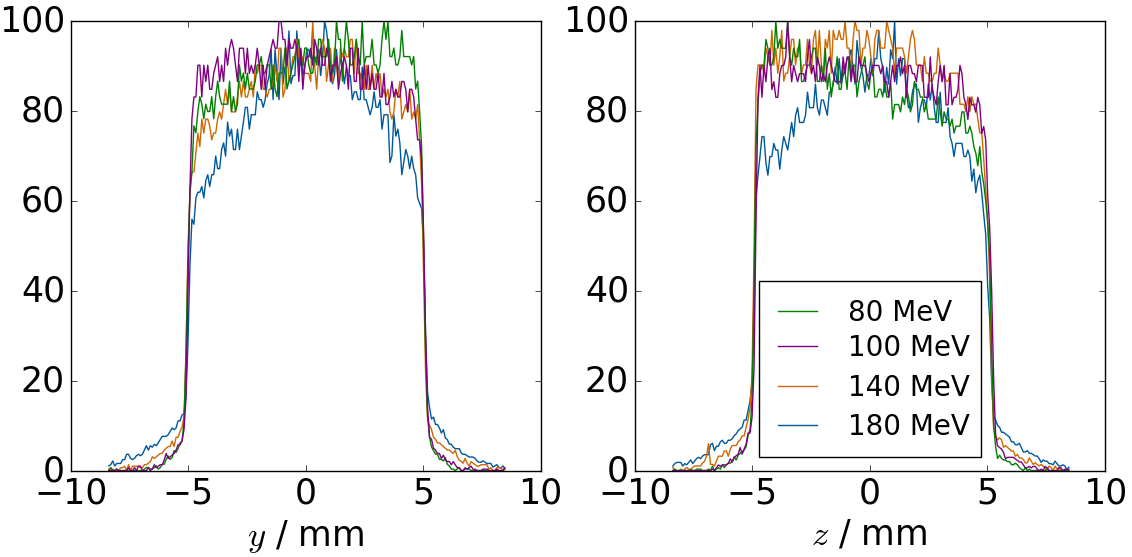} \label{fig:1DBeamProfile} }}
    \caption{\label{fig:profiles} Measured beam profiles at the distal edge of the collimators. (a) Exemplary 2D beam profile for \SI{180}{MeV} and (b) central line profiles for different energies between 80 and \SI{180}{MeV}. Profiles were measured with an EBT3 film dosimeter attached to the back of the last collimator, perpendicular to the beam.}
\end{figure}

In order to reduce the particle fluence on the magnets and the yoke and thus to prevent radiation damage, the beam was collimated by two brass collimators with a circular aperture of \SI{10}{mm} diameter. The collimators had a cylindrical shape with an outer radius of \SI{9}{cm} and a combined thickness of \SI{6.6}{cm}. This thickness allows to fully stop peripheral protons for all used energies up to \SI{180}{MeV}. The distance between the beam exit window and the surface of the first collimator was \SI{1.63}{m}, and that between the two collimators was \SI{2}{cm}. The large distance between the beam exit and the collimators was chosen to improve the lateral beam homogeneity by increased scattering in air. The collimators were placed in the beam line such that a symmetric beam profile was achieved behind the collimators (see Extended Data, Figure \ref{fig:profiles}).

The collimated beam travelled \SI{22}{cm} through air before it impinged on the phantom, which was placed centrally in the beam's isocentre inside a transverse magnetic field. The phantom comprised two slabs of polymethyl methacrylate (PMMA) and measured \mbox{\(x_\mathrm{ph} \times y_\mathrm{ph} \times z_\mathrm{ph} =\) \SI{30 x 15 x 3}{} \SI{}{\cubic\centi\metre}}. The density of the PMMA was \mbox{\((1.186 \pm 0.002) \SI{}{g\,cm^{-3}}\)} as measured by density determination scales. The phantom was mounted into the air gap of the magnet assembly by an L-shaped holder.

In order to measure planar dose distributions in the isocentre plane, a film detector was placed horizontally in the phantom. The film was bevelled by \(\alpha = \SI{1}{\degree}\) to reduce the dependence of the dose distribution on the film material and possible air gaps between phantom and film. An angle of \(\alpha = \SI{1}{\degree}\) was chosen as it tilted the measurement plane minimally out of the central beam plane whilst suppressing these effects as good as larger angles \cite{Zhao2010}. With this angle, the inclination lead to differences in \(x\) below \SI{0.02}{mm}. However, we note that this tilt makes the resulting depth-dose curves not directly comparable to those which would be obtained with an ionization chamber moving along the central beam axis due to the laterally sloping beam profile. The air gap between the phantom and the film was minimized by four plastic screws pressing the slabs together.

The magnet assembly consists of two magnet poles with a size of \mbox{\(x_\mathrm{pole} \times y_\mathrm{pole} \times z_\mathrm{pole} =\) \SI{20 x 15 x 5.9}{} \SI{}{\cubic\centi\metre}} each and a density of \SI{7.6}{\gram\per\cubic\centi\metre} \cite{Vacodym2014}. The two poles made up of \chem{Nd_2Fe_{14}B} are mounted on a C-shaped yoke, mounted on a mobile table. The air gap between the poles has a height of \SI{4}{cm}. 

As opposed to a previous study for photons \cite{Reynoso2016}, no systematic differences in absolute dose were found between the film measurements with and without magnetic field, which may be caused by the fact that film dosimeters were oriented perpendicular to the magnetic field in this study. 

\subsection{Film handling and evaluation} \label{cha:films}

EBT3 Gafchromic films (Ashland, Covington, USA; lot 05201501) cut to \mbox{(\SI{20 x 15}{}) \SI{}{\centi\metre^2}} were used to measure the planar dose distribution of the beam. They were scanned with a flat-bed document scanner (Expression 11000XL, Epson America, Long Beach, CA) in landscape orientation, transmission mode, 24-bit color mode, and a resolution of \SI{300}{dpi}. 

Scanned images were analysed using in-house developed software written in the Python programming language (Python Software Foundation). The net optical density (netOD) was computed from the red channel pixel intensity \(I\) and background intensity \(I_0\) for each pixel \(i\) as
\begin{equation}
\mathrm{netOD}_i = \mathrm{log}_{10}\frac{I_0}{I_i} \ , 
\end{equation}
and converted to dose using the calibration
\begin{equation}
D_i = k_1 (\mathrm{netOD}_i) + k_2 (\mathrm{netOD}_i) ^{k_3} 
\end{equation}
with \(k_1 = 8.36\), \(k_2 = 10.71\) and \(k_3 = 1.84\). The calibration had been performed in analogy to a previous study \cite{Zeil2009}. The first 30 pixels (\SI{2.55}{mm}) around film cutting edges were omitted.

Pixels were converted to \(x\)-\(y\)-coordinates as follows. Pixels were scaled with the image resolution, yielding a distribution on the film plane in \(x'\) and \(y'\). This coordinate system is tilted by \(\alpha = \SI{1}{\degree}\) around the \(y\)-axis relative to the system defined above, which yields \(x = \cos(\alpha) x'\) and \(y = y'\). Note that \(x - x' < \SI{0.02}{mm}\) within the film. The pinholes on the scanned images, whose \(x\)-\(y\)-positions are given by the manufacturer, were automatically detected on the scanned images and hence defined a translation and rotation of the dose distribution.

\subsection{Out-of-plane beam deflection} \label{cha:outofplane}

\begin{figure}
\begin{center}
 \subfloat[]{{\includegraphics[width=.48\textwidth]{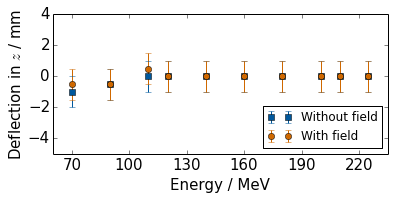}\label{fig:LynxZ} }} 
 \subfloat[]{{\includegraphics[width=.50\textwidth]{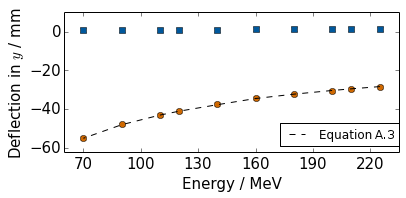}\label{fig:LxnxY} }}
 \caption{\label{fig:Lynx} Measured beam deflection in \(z\)- (a) and \(y\)-direction (b) on the scintillation detector after transmission through the magnet air gap without phantom. Positions of beam profile maxima with (orange) and without (blue) the magnet in place are displayed.}
\end{center}
\end{figure}

A transmission irradiation experiment was performed to ensure that the beam was not deflected out of the  film (\(x\)-\(y\)-) plane by magnetic field components pointing along the \(y\)-axis. For this purpose, the phantom was removed and a pixelated scintillation detector with a resolution of \SI{0.5}{mm} (Lynx, IBA Dosimetry, Schwarzenbruck, Germany) was placed behind the magnet assembly in a distance of \SI{24}{cm}. As depicted in Extended Data Figure \ref{fig:LynxZ}, the out-of-plane deflection due to the magnetic field was within measurement uncertainty.

The deflection in \(y\)-direction observed on the scintillation detector \(y_\mathrm{Lynx}\) is depicted in  Extended Data Figure \ref{fig:LxnxY}. We note that, assuming a uniform magnetic flux density \(B = \SI{0.95}{T}\) between the magnet poles and no magnetic field outside, \(y_\mathrm{Lynx}\) can be easily approximated by trigonometric considerations from the length of the magnet poles \(x_\mathrm{pole}\) and the distance between the magnet and the scintillation device \(d_\mathrm{air}\) as
\begin{equation}
y_\mathrm{Lynx} = r - \sqrt{r^2-x_\mathrm{pole}^2}+\frac{d_\mathrm{air}}{\sqrt{r^2-d_\mathrm{air}}} \, .
\end{equation}
Neglecting relativistic effects and beam energy loss in air, the radius of the beam path inside the magnetic field is \mbox{\(r \approx 14.4 \frac{\sqrt{E_0}}{B_0} \, \SI{}{T \, cm \, (MeV)^{-\frac{1}{2}}}\)}, and a function of the initial proton energy \(E_0\) \cite{Hartman2015}. Values for \(y_\mathrm{Lynx}\) obtained with this approximation for a nominal distance between magnet and detector screen of \(d_\mathrm{air} = \SI{26}{cm}\) show excellent agreement with the measured positions (see Extended Data, Figure \ref{fig:LxnxY}). The difference of \(d_\mathrm{air}\) to the measured distance between the magnet and the detector can be ascribed to the distance between the active scintillation layer and the detector's surface and the non-uniformity of the magnetic field.

\subsection{Uncertainty estimation} \label{cha:uncertainty}

The influence of the measurement uncertainties on the Bragg peak lateral deflection \(\Delta y_{80}\) and range retraction \(\Delta R_{80}\) was assessed as follows (see Table \ref{tab:expuncertainty}).

Statistical uncertainties such as the robustness of the \(\Delta R_{80}\)- and \(\Delta y_{80}\)-determination, and possible remaining displacement of the film relative to the phantom and to the scanner have been assessed by repeating the measurement for \SI{180}{MeV} three times. Mean absolute deviations were used as statistical uncertainties.

Film-to-film variations and the film calibration are subject to percentual dose uncertainties. The calibration has a relative dose uncertainty of \SI{5}{\%} \cite{Zeil2009}. Dose uncertainties due to film and scanner variations add up to \SI{0.5}{\%} \cite{Sorriaux2012}. Being proportional to the dose, these uncertainties were considered to have negligible influence on \(\Delta R_{80}\) and \(\Delta y_{80}\), which only depend on the relative dose difference between the measurement with and without magnetic field.

The influence of the LET dependence of the film response was assessed by applying a polynomial correction function \cite{Zhao2010} to the depth-dose curves. Taking into account the film-, energy-, dose- and possible setup-dependence of this method, this correction is not directly transferrable to this study but serves as a measure to estimate the influence of this effect. Thus, we estimate the uncertainty caused by this effect to be as high as the difference in \(\Delta R_{80}\) and \(\Delta y_{80}\) between the corrected and uncorrected data.

The setup was aligned using the in-room laser system, rulers and a spirit level. Systematic setup uncertainties are related to translational and rotational degrees of freedom.

A rotation of the film dosimeter relative to the beam axis has a direct impact on the mesured peak deflection. The alignment of the axes of the beam and the laser positioning system was uncertain to within \SI{0.1}{^o}. The consequence of this angle on Bragg peak retraction and deflection have been assessed by purposefully rotating the dose distribution by this angle.

Rotation of the magnet relative to the laser and of the phantom relative to the magnet were assessed by repeating the setup and irradiation procedure. For this analysis, the beam trajectory was determined for each depth as the mean of the two lateral positions receiving \SI{10}{\%} of the maximum dose. As no systematic dependence on the beam energy was observed in the deviations between the two experiments, mean absolute deviations were used as an estimate for this uncertainty. Rotations of the film relative to the phantom are statistical and therefore already covered.

Translational uncertainties in \(x\)-direction were considered to be negligible, as they mostly lead to uncertainties in the amount of air traversed. An exception to this is the uncertainty of the pin positions (see Supporting \ref{sec:expsetup}), which was below \SI{10}{\micro\meter} and therefore negligible. Translational uncertainties in \(y\)-direction were suppressed by setting the mean of the \(y\)-coordinate of the trajectory in the first 10 pixels to zero. Proton beam deflection in the magnetic fringe field outside of the magnet assembly was considered negligible.

Another factor of uncertainty was introduced by the initial spatial achromaticity of the beam, i.e. a \(y\)-\(z\)-dependent energy spectrum, caused by the beam line design. This resulted in a kink towards the end of the trajectory, which lead to an overestimation of the beam deflection. This effect was observed both with and without magnetic field. The kink is about \SI{3}{mm} in \(x\)- and \SI{0.2}{mm} in \(y\)-direction.

Relevant sources of uncertainty are summarized and quantified in Supporting Table \ref{tab:expuncertainty}. The total uncertainty was estimated as the square root of the quadratic sum of the individual contributions.

\begin{figure}
\begin{center}
 \subfloat[]{{\includegraphics[width=.59\textwidth]{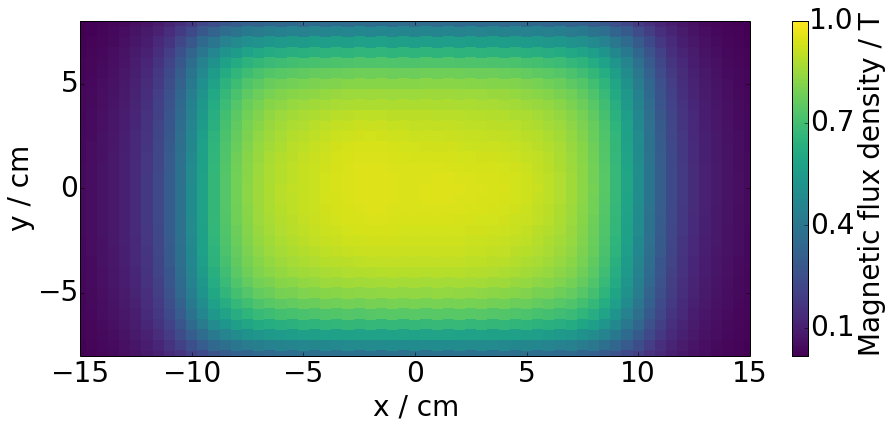}\label{fig:field2D} }}
 \subfloat[]{{\includegraphics[width=.27\textwidth]{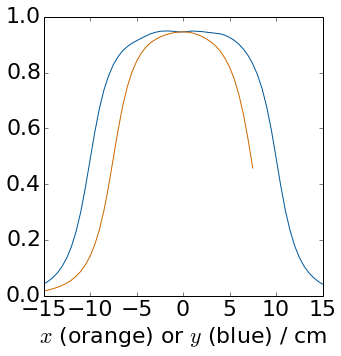}\label{fig:field1D} }}
 \caption{\label{fig:field} Central plane of the measured magnetic field (a) and central field profiles (b) along the \(x\)-(blue) and \(y\)-(orange) axes. Values for \(y > \SI{7.5}{cm}\) are not displayed due to the presence of the magnet yoke.}
\end{center}
\end{figure}

\begin{table}[htb]
\setlength{\tabcolsep}{1.0mm}
\caption{Uncertainties of the measured lateral deflection \(\Delta y_{80}\) and range retraction \(\Delta R_{80}\) of the Bragg peak. Lower and upper limits are given for energy-dependent effects, where lower energies are related to smaller uncertainties.}
\label{tab:expuncertainty}
\begin{center}
\begin{minipage}{.6\textwidth}
\small
\begin{tabular}{l|c|c}
		\hline
		\hline 
Source of uncertainty & \(\Delta R_{80}\)/mm & \(\Delta y_{80}\)/mm\\ 
		\hline
		\hline 
\textbf{Statistical}	& 0.4 & 0.2 \\
		\hline
\textbf{Sytematic} & 0.3 ... 0.4 & 0.4 \\
Thereof: & & \\
\hspace*{3mm} LET dependence & 0.0 ... 0.2 & 0.0 ... 0.2\\
\hspace*{3mm} Setup axis alignment & 0.3 	 & 0.3 \\
\hspace*{3mm} Beam achromaticity & $<$ 0.01 & 0.2\\ 
		\hline

\textbf{Total}	   & 0.5 & 0.4 ... 0.5 \\
		\hline
		\hline
\end{tabular}
\normalsize
\end{minipage}
\end{center}
\end{table}

\section{Monte Carlo Simulations}

Simulations were conducted on our high-performance computing cluster. Data analysis was performed using the Python programming language (Python Software Foundation. Python Language Reference, version 3.4.3, www.python.org).

\subsection{Beam model definition} \label{sec:beammodel}

As the proton beam is not mono-energetic, beam energy distributions corresponding to the facility's nominal beam energies \(E_\mathrm{nom}\) needed to be determined. For this purpose, in-water depth-dose curves were measured for 24 energies between \SI{100}{MeV} and \SI{215}{MeV} using a multi-layer ionization chamber with \SI{6}{cm} radius and an effective depth resolution of \SI{1}{mm} (Giraffe, IBA Dosimetry, Schwarzenbruck, Germany) \cite{Wohlfahrt2017}. An analytical approximation of the Bragg curve \cite{Bortfeld1997} was fitted to the measured data, yielding the proton range \(R_{80}\) and its energy spread \(\sigma_E\). Ranges \(R_{80}\) were converted to mean initial proton energies \(E_0\) using tabulated stopping power data \cite{PSTAR}. \(E_0\) and \(\sigma_E\) are given in Table \ref{tab:energies} for all studied energies.

The obtained energies \(E_0\) exceeded the nominal energies \(E_\mathrm{nom}\) by \SI{1}{MeV} to \SI{2}{MeV}, which can be explained by the fact that the beam line is calibrated to produce certain beam ranges, whereas the nominal energies serve as labels. The measured energy spread \(\sigma_E\) ranged between \SI{0.8}{MeV} and \SI{1.2}{MeV}. \(E_0\) and \(\sigma_E\) were used to define the beam for the Monte Carlo simulations.

If not stated differently, energies mentioned throughout this work refer to the nominal energies.

\begin{table}[h]
\centering
\setlength{\tabcolsep}{1.45mm}
\caption{Beam energy \(E_0\) and energy spread \(\sigma_E\) obtained from fit to depth-dose measurements for nominal energies \(E_\mathrm{nom}\). All quantities are in MeV.}
\label{tab:energies}
\begin{center}
\begin{minipage}{0.8\textwidth}
\small
\begin{tabular}{|ccc|}
\hline 
\(E_\mathrm{nom}\) & \(E_0\) & \(\sigma_E\) \\ 
\hline 
100 & 101.8 & 0.8 \\ 
 
105 & 107.1 & 0.9 \\ 
 
110 & 112.2 & 0.9 \\ 
 
115 & 117.2 & 1.0 \\ 
 
120 & 121.9 & 1.0 \\ 
 
125 & 126.7 & 1.1 \\ 
 
130 & 131.9 & 1.1 \\ 

135 & 137.1 & 1.1 \\ 
\hline 
\end{tabular}
\begin{tabular}{|ccc|}
\hline 
\(E_\mathrm{nom}\) & \(E_0\) & \(\sigma_E\) \\ 
\hline 

140 & 141.8 & 1.2 \\ 

145 & 146.8 & 1.2 \\ 

150 & 151.6 & 1.2 \\ 

155 & 156.9 & 1.2 \\ 

160 & 161.9 & 1.2 \\ 

165 & 166.9 & 1.2 \\ 

170 & 171.7 & 1.3 \\ 

175 & 176.6 & 1.3 \\ 
\hline 
\end{tabular}
\begin{tabular}{|ccc|}
\hline 
\(E_\mathrm{nom}\) & \(E_0\) & \(\sigma_E\) \\ 
\hline 

180 & 181.8 & 1.3 \\ 

185 & 186.8 & 1.3 \\ 

190 & 192.0 & 1.3 \\ 

195 & 196.9 & 1.3 \\ 

200 & 201.7 & 1.3 \\ 

205 & 206.9 & 1.3 \\ 

210 & 211.9 & 1.3 \\ 

215 & 216.9 & 1.2 \\ 
\hline 
\end{tabular}

\normalsize
\end{minipage}
\end{center}
\end{table}

To acquire a realistic model of the source, 2D beam profiles were measured close to the beam exit window (in a distance of \SI{8}{cm}) using a the pixelated scintillation detector for seven energies between \SI{100}{MeV} and \SI{215}{MeV}. A fit of a two-dimensional Gaussian function to the data yielded a mean lateral spread of \({\sigma_0}^{yz}=(4.7 \pm 0.7) \SI{}{mm}\) with a small trend of decreasing \({\sigma_0}^{yz}\) with increasing energy and a slightly higher spread in vertical (\(z\)-) direction than in horizontal (\(y\)-) direction.

Varying the simulated initial beam spread \({\sigma_0}^{yz}\) between 0 and \SI{8}{mm} showed no influence on the simulated beam spread at the phantom entrance larger than \SI{0.1}{mm}. Thus, the beam shape at the phantom entrance is mainly determined by scattering in the \SI{202}{cm} long air gap between the beam exit and the phantom and by stopping in the collimators, and not by the initial beam profile. For simplicity, the initial beam profile was therefore modelled as an energy-independent Gaussian with a fixed unilateral spread of \({\sigma_0}^{yz}=\SI{4.7}{mm}\).\\

Different models of the physical processes involved are available within Geant4, with a set of such process models being called a physics list. A number of different recommendations for physics lists for proton therapy have been made in previous studies and by the Geant4 and GATE collaborations: \code{QGSP\_BIC\_HP\_EMY} \cite{Almhagen2015}, \code{QGSP\_BIC\_HP\_EMZ}, \code{QGSP\_BERT\_HP\_EMY} and \code{QGSP\_BERT\_HP\_EMZ},  \cite{Geant4Reference},  \code{HADRONTHERAPY\_1} and \code{HADRONTHERAPY\_2} \cite{Geant4HADRO}, and \code{QBBC\_EMY} and \code{QBBC\_EMZ} \cite{Grevillot2010,GATEphysics}. These physics lists differ mainly in the representation of hadronic and electromagnetic interactions, as well as low-energy neutron interactions and radioactive decay models. A detailed description of the relevant models and physics list can be found in \cite{Jarlskog2008}. Since there is no clear consensus within these recommendations for a single superior physics list, all mentioned physics lists were included and tested in this study.

The ionization potential of water was set to \SI{78}{eV}, which is the default value for Geant4 version 10 and higher, and is consistent with the revised ICRU report 73 \cite{ICRU73}. Note that this value differs from the one used in tabulated stopping-power data \cite{PSTAR} by \SI{3}{eV}. Parameters defining the resolution of the simulation, i.e. secondary particle production threshold, maximum step length and dose scorer resolution, were defined by a convergence study, meaning that they were reduced until no change in the observables was found within a required precision. Observables were the proton range and energy spread with a required precision of \SI{0.1}{mm} and \SI{0.1}{MeV}, respectively. For increased simulation efficiency, the secondary particle production threshold was determined independently inside and outside of the phantom (the latter being referred to as \textit{world}).

The \code{RANMAR} random number generator was chosen, as it provides \SI{9e8}{} disjoint sequences with a length of \( \SI{e30}{} \) numbers \cite{James1990}. The magnetic field was implemented using a corrected version \cite{bug} of the Geant4 class \code{HadrontherapyMagneticField3D} given in the \code{HADRONTHERAPY} application example of Geant4.

\subsection{Beam model verification} \label{sec:validation}

As a verification for the thus defined beam model, depth-dose curves were scored in a water phantom (size \mbox{\(x_\mathrm{wp} \times y_\mathrm{wp} \times z_\mathrm{wp} =\) \SI{35 x 12 x 12}{} \SI{}{\cubic\centi\metre}}). For 24 energies, the range \(R_0\) and energy spread \(\sigma_E\) were obtained by fitting the analytical Bragg curve \cite{Bortfeld1997} to these depth-dose curves (as in \ref{sec:beammodel}). These were then compared to those obtained from the experimental data.

Furthermore, two-dimensional beam profiles were measured at the phantom entrance position behind the collimators using the pixelated scintillation detector. Profiles were scored in the simulation at the same position. Spatial spreads at the phantom entrance position \({\sigma_\mathrm{ph}}^{yz}\) were extracted by fitting a bivariate Gaussian function to the obtained experimental and simulated planar dose profiles.

The physics model \code{QGSP\_BERT\_HP\_EMZ} showed the smallest deviation to the experimental data and was thus chosen for the simulation, although differences between the studied models in range and energy spread were smaller than \SI{0.2}{mm} and \SI{0.05}{MeV}, respectively. The minimum required production threshold for the \textit{world} was found to be \SI{5}{mm} and the minimum required production threshold for the phantom, step limiter and scorer resolution were found to be \SI{0.1}{mm}. This corresponds well with values found in previous studies \cite{Grevillot2010,Kurosu2014}. 

Mean differences between the model and reference measurements of depth-dose curves and lateral beam profiles in water without magnetic field (see \ref{sec:validation}), were \SI{0.2}{mm} in range (max. \SI{0.5}{mm}), \SI{0.23}{MeV} in energy spread (max. \SI{0.35}{MeV}) and \SI{0.1}{mm} in spatial spread (max. \SI{0.2}{mm}). The beam model was thus accepted for the simulations. 

Dose was scored in the PMMA phantom at the position of the film detector, on a \mbox{\SI{20 x 15}{} \SI{}{\square\centi\metre}} plane with a thickness of \SI{30}{\micro\meter} corresponding to that of the active layer of a Gafchromic EBT3 film dosimeter (Ashland, Covington, USA). 

We note that the parameters defining the spatial and energy distributions were directly implemented as measured, and not adjusted iteratively to improve the match between simulated and measured data. This allowed an unbiased comparison of the different physics lists.

For the deflection study (see main manuscript), the convergence study for the simulation resolution parameters was repeated with the range and the Bragg peak deflection as observables and a required accuracy of \SI{0.1}{mm}. The same parameters were found as for the model building (\SI{5}{mm} for the \textit{world} production threshold, and \SI{0.1}{mm} for the phantom production threshold, step limiter and scorer resolution). The film dosimeter was not explicitly simulated to avoid boundary artefacts.

The most relevant simulation parameters are summarized in table \ref{tab:parameters}.

\begin{table}[htb]
\setlength{\tabcolsep}{1.0mm}
\caption{Geant4 simulation parameters.}
\label{tab:parameters}
\begin{center}
\begin{minipage}{.6\textwidth}
\small
\begin{tabular}{l c}
		\hline
		\hline 
Geant4 version & 10.2.p02 \\
Physics list & \code{QGSP\_BERT\_HP\_EMZ}\\ 
Production threshold in \textit{world}& \SI{5}{mm} \\
Production threshold in phantom & \SI{0.1}{mm}\\
Step limiter & \SI{0.1}{mm}\\
Scorer resolution &  \SI{0.1}{mm}\\
Water mean excitation energy & \SI{78}{eV}\\
Random number generator &\code{RANMAR} \\ 
		\hline
		\hline
\end{tabular}
\normalsize
\end{minipage}
\end{center}
\end{table}

We note that in the depth-dose curves (see figure \ref{fig:Depthdose}) of the \SI{180}{MeV} beams, a small peak is observed at a depth of about \SI{2.5}{cm} both with and without magnetic field. The peak was not reproduced when the two collimators were replaced by one collimator of their combined thickness. Furthermore, the proton energy spectrum at the phantom entrance showed a steep maximum around \SI{180}{MeV} and a small maximum around \SI{60}{MeV}. Thus, a probable explanation for the small peak are low energy protons having traversed only one of the two collimators. 

\subsection{Magnetic field model}

A model of the magnetic field was generated using finite-element modelling (COMSOL Multiphysics, COMSOL AB, Stockholm, Sweden). The geometry of the magnet assembly was implemented as given by the manufacturer (Vacuumschmelze GmbH, Hanau, Germany) and surrounded by the pre-defined material 'Air'. The remanent field strength of the magnetic poles was set to \(B_r = \SI{1.37}{T}\), as specified for the VACODYM 764 TP material by the manufacturer \cite{Vacodym2014}. The hysteresis curve \(B(H)\) for the yoke was inherited from the COMSOL material library using the predefined material 'Soft Iron (without losses)'. The model was validated comprehensively using an automated magnetometry setup comprising a robotic positioning device and a Hall probe (MMTB-6J04-VG, Lake Shore Cryotronics, Westerville, USA) connected to a Gaussmeter (Model 421, Lake Shore Cryotronics). Predicted and measured magnetic flux densities agreed within \SI{2}{\%} \cite{Gantz2017}.

In air, the field of the magnet is mainly limited to the gap between its two poles, and has a maximum magnetic flux density of \SI{0.95}{T} (see Figure \ref{fig:field}). Its main component \(B_z\) is aligned parallel to the \(z\)-axis, which leads to a deflection of the proton beam in (positive) \(y\)-direction. 

A look-up-table (LUT) of the magnetic field was exported from the finite-element model with a grid spacing of \SI{5}{mm} in a volume of \mbox{\(x_\mathrm{B} \times y_\mathrm{B} \times z_\mathrm{B} =\) \SI{70 x 40 x 20}{} \SI{}{\cubic\centi\metre}}. The LUT covered the three magnetic field components in the volume of \(x \in [\SI{-50}{cm},\SI{20}{cm}]\) and \(y \in [\SI{-20}{cm},\SI{20}{cm}]\) and \(z \in [\SI{-10}{cm},\SI{10}{cm}]\).

\subsection{Uncertainty estimation} \label{sec:simuncertainty}

To assess the influence of the different uncertainties of the simulations, a number of separate Monte Carlo simulations were performed while varying the input parameters. For each of these uncertainty simulations, the Bragg peak deflection \(y_{80}\) and retraction \(\Delta R_{80}\) were extracted and compared to the original simulations. All uncertainties were calculated for the lowest (\SI{80}{MeV}) and highest (\SI{180}{MeV}) energy used and interpolated linearly for intermediate energies.

Relevant systematic uncertainties of the input parameters are those of the initial proton energy \(E_0\), of the density of the PMMA phantom \(\rho\) and of the main magnetic field component \(B_z\). 

The uncertainty of the range measurements performed for \ref{sec:beammodel} amounts to \SI{0.25}{mm}. Using tabulated stopping-power data \cite{PSTAR}, this corresponds to an uncertainty of the beam energy \(E_0\) of \(\Delta E_0 (\SI{80}{MeV}) = \SI{0.25}{MeV}\) to \(\Delta E_0 (\SI{180}{MeV}) = \SI{0.15}{MeV}\). The absolute uncertainty of the PMMA density amounts to \(\Delta\rho= \SI{0.002}{g\,cm^{-3}}\), and that of the simulated magnetic field map to \(\Delta B_z = \SI{0.02}{T}\). 

Each of these parameters was separately increased by its uncertainty (i.e. \(E_0+\Delta E\), \(\rho + \Delta\rho\) and \(B_z + \Delta B_z\)) in an additional simulation to assess its influence on \(y_{80}\) and \(\Delta R_{80}\). It was assumed that the influence of the parameter uncertainties was equally large in positive and negative direction.

Furthermore, there is an influence of systematic uncertainties which are inherent to the Monte Carlo simulation, for example in the physical process models and the mean excitation energy applied. As an estimate for this uncertainty, the range differences of the model to the reference measurements (\ref{sec:validation}) of \(\Delta R = \SI{0.2}{mm}\) was used. The influence of this range uncertainty on \(y_{80}\) and \(\Delta R_{80}\), was assessed by varying the initial energy accordingly, i.e. with \(\Delta E_0^R (\SI{80}{MeV}) = \SI{0.20}{MeV}\) to \(\Delta E_0^R (\SI{180}{MeV}) = \SI{0.12}{MeV}\). 

The statistical uncertainty of the simulation was estimated by repeating the original simulation five times with different random number seeds and calculating the standard deviation of \(y_{80}\) and \(\Delta R_{80}\). It was found that all simulations would be performed with \SI{2.048e7}{} primary particles each to acquire a statistical uncertainty of the Bragg peak position of \SI{0.1}{mm} or smaller.

The total uncertainty was estimated as the square root of the quadratic sum of the individual contributions.

\begin{table}[htb]
\caption{Uncertainties of the predicted lateral deflection \(\Delta y_{80}\) and range retraction \(\Delta R_{80}\) of the Bragg peak. Lower and upper limits are given for energy-dependent effects, where lower energies are related to smaller uncertainties.}
\label{tab:simuncertainty}
\begin{center}
\begin{minipage}{.6\textwidth}
\small
\begin{tabular}{l@{\hskip 8mm}c@{\hskip 8mm}c}
		\hline
		\hline 
Source of uncertainty & \(\Delta R_{80}\)/mm & \(y_{80}\)/mm\\ 
		\hline
		\hline 
\textbf{Systematic} & 0.1 ... 0.2 & 0.1 ... 0.3 \\
Thereof: & & \\
\hspace*{3mm} Beam energy & 0.1 & 0.0 ... 0.1 \\
\hspace*{3mm} Phantom density & 0.0 & 0.1 \\
\hspace*{3mm} Magnetic field & 0.0 ... 0.1 & 0.0 ... 0.2 \\ 
\hspace*{3mm} Physics modelling & 0.0 ... 0.1 & 0.0 ... 0.1 \\
		\hline
\textbf{Statistical}	& 0.0 ... 0.1 & 0.0 ... 0.1 \\
		\hline

\textbf{Total}	   & 0.1 ... 0.2 & 0.1 ... 0.3 \\
		\hline
		\hline
\end{tabular}
\normalsize
\end{minipage}
\end{center}
\end{table}

Almost none of the individual uncertainty contributions exceeded the resolution of the dose scoring grid (\SI{0.1}{mm}). The only exception to this is the uncertainty caused by the magnetic field map, with an uncertainty of \SI{0.2}{mm} for the Bragg peak deflection at \SI{180}{MeV}. With a relative uncertainty of \SI{2}{\%}, the magnetic field map is the largest factor of uncertainty to this study. The fact that it causes an uncertainty of approximately \SI{2}{\%} in the Bragg peak deflection is consistent with the proportionality of the two as found in a previous study \cite{Wolf2012}.

Absolute uncertainties tend to increase with increasing energy, whereas relative uncertainties tend to decrease. Note that, as the systematic uncertainties were gained in analogous simulations as the main simulations, they are themselves subject to a statistical uncertainty of up to \SI{0.1}{mm}.

\section{Depth dose and trajectory reconstruction} \label{sec:integration}
Two-dimensional dose distributions \(D(x,y)\) were obtained from the film plane. From these, the lateral \mbox{(\(y\)-)} deflection and longitudinal (\(x\)-) retraction of the Bragg peak due to the magnetic field were extracted as follows. For each depth \(x\), a univariate Gaussian function was fitted to the lateral beam profile and the maximum position \(y_\mathrm{T}\) was extracted. This yielded a function for the lateral maximum dependent on depth, the beam trajectory \(y_\mathrm{T}(x)\). To mimic a depth-dose curve measured by an ionization chamber of \SI{8}{cm} diameter, for each depth the predicted dose was integrated radially around \(y_\mathrm{T}\) with an integration radius of \(r=\SI{4}{cm}\). This resulted in a function for the integrated lateral dose dependent on depth, the integral depth-dose curve \[\mathit{IDD}(x)=\int_0^{2\pi}\int_0^r D(r',\phi)r'dr'd\phi=\pi \int_{-r}^{r} D(x,y) |y-y_\mathrm{T}(x)| dy \] with \(r'\) and \(\phi\) the polar coordinates. Both \(y_\mathrm{T}(x)\) and \(\mathit{IDD}(x)\) were smoothed to reduce statistical noise using a univariate spline fit. The range \(R_{80}\) was calculated from \(\mathit{IDD}(x)\) as the depth of the beam at the \SI{80}{\%} distal end of the \(\mathit{IDD}\) maximum. The longitudinal Bragg peak retraction \(\Delta R_{80}\) was defined as the difference between \(R_{80}\) with and without magnetic field. The lateral Bragg peak deflection \(y_{80}\) was determined from the trajectory obtained with magnetic field as \(y_\mathrm{T}(R_{80})\).

\section{Bragg peak displacement}

Measured and simulated Bragg peak deflection and retraction values and their uncertainties are summarized in Supporting Table \ref{tab:deflection}.

\begin{table}[htb]
\caption{Bragg peak displacement caused by the magnetic field in lateral (\(\Delta y_{80}\)) and longitudinal (\(\Delta R_{80}\)) direction for different proton energies. Positive retraction values indicate that the range with magnetic field is reduced compared to the reference range without field.}
\label{tab:deflection}
\begin{center}
\begin{minipage}{0.6\textwidth}
\begin{tabular}{c|c|c|c|c}
		\hline
		\hline 
		& \multicolumn{2}{c|}{Measurement (EBT3)}& \multicolumn{2}{c}{Simulation (Geant4)}\\
\(E_\mathrm{nom}\) / MeV & \(\Delta y_{80}\) / mm & \(\Delta R_{80}\) / mm & \(\Delta y_{80}\) / mm & \(\Delta R_{80}\) / mm\\ 
\hline 
80  & \(1.0 \pm 0.4\) & \(0.0 \pm 0.5\)& \(1.0 \pm 0.1\) & \(0.1 \pm 0.1\)\\
100 & \(1.7 \pm 0.4\)& \(-0.5 \pm 0.5\)& \(1.7 \pm 0.1\)& \(0.0 \pm 0.1\)\\
120 & \(3.2 \pm 0.4\)& \(0.3 \pm 0.5\) & \(2.9 \pm 0.2\)& \(0.1 \pm 0.1\)\\
140 & \(4.3 \pm 0.4\)& \(-0.2 \pm 0.5\)& \(4.6 \pm 0.2\)& \(0.2 \pm 0.2\)\\
160 & \(6.9 \pm 0.4\)& \(-0.2 \pm 0.5\)& \(6.7 \pm 0.3\)& \(0.2 \pm 0.2\)\\
180 & \(10.1 \pm 0.5\)& \(0.4 \pm 0.5\)& \(9.3 \pm 0.3\)& \(0.3 \pm 0.2\)\\ 
		\hline
		\hline
\end{tabular}
\end{minipage}
\end{center}
\end{table}

\end{document}